# COMPARISON OF THE RESEARCH EFFECTIVENESS OF CHEMISTRY NOBELISTS AND FIELDS MEDALIST MATHEMATICIANS WITH GOOGLE SCHOLAR: THE YULE-SIMON MODEL

By

Stephen J. Bensman,
LSU Libraries
Louisiana State University
Baton Rouge, LA 70803 USA
E-mail: notsjb@lsu.edu

Lawrence J. Smolinsky
Department of Mathematics
Louisiana State University
Baton Rouge, LA 70803 USA
E-mail: smolinsk@math.lsu.edu

Daniel S. Sage
Department of Mathematics
Louisiana State University
Baton Rouge, LA 70803 USA
E-mail: sage@math.lsu.edu

**Abstract**

This paper presents a test of the validity of using Google Scholar (GS) to assess researchers' publication impacts. It does this by using the Yule-Simon model to compare the research effectiveness of chemistry Nobelists and Fields medalist mathematicians as measured by GS inlinks. It finds that GS delivers results that are consistent with the structure of the disciplines and the nature of the prizes. Chemistry is a science, and the Nobel prize is awarded at the end of a chemist's career. Mathematics is a fragmented discipline, more philosophical than scientific in many respects, and the Fields medal is awarded at the beginning of a mathematician's career. Due to these factors, the Yule-Simon better approximates the inlink distributions to the works of the chemists than to those of the mathematicians, which manifest a more random pattern.

## Introduction

This paper presents a test of the validity of using Google Scholar (GS) to assess researchers' publication impacts. The test is based upon the Yule-Simon distribution, and it is applied to Nobel laureates in chemistry and Fields medalists in mathematics. Both the disciplines and the awards are structurally different from each other. The Yule-Simon distribution is treated as a mathematical construct to assess the impact of the laureates and medalists as measured by their publications and GS inlinks[1]. If the differences in impact found by the test matches the structural differences between the disciplines and the awards, then the test can be considered a validation of not only GS but also of the Yule-Simon model as a way of measuring publication impact.

The validation of GS and the Yule-Simon model is particularly important and interconnected. As a result of the information revolution, scientific and scholarly documents have moved from paper to the World Wide Web, and the connection between them has changed from the citation to the hyperlink, making webometrics an important component of scientometrics. GS is currently the most useful route for retrieving scientific documents from the Web and quantifying the hyperlinks between them. However, it is questionable whether the Google search engine can separate the wheat from the chaff dominating the Web in order to form the relevant document sets necessary for assessing researchers' publication impacts.

The Yule-Simon model can play an important role in validating GS. It was originally derived by George Udny Yule, a British biometrician and pioneer of modern inferential statistics, to model the development of species distributions over time. Yule's initial model was further developed by Herbert Simon, a Nobel laureate in economics, who expanded its application to socioeconomic and informetric data. In this form, the Yule-Simon distribution has been found to be a feasible model not only of empirical informetric laws, but also of the distributions dominating the Web. Its method of measuring the development of a distribution to completion over time allows the creation of a statistical technique to measure how well and how far researchers' works have been incorporated into the knowledge corpus of their disciplines. This technique is presented and explained in this chapter so that others can use it to analyze the creation and codification of knowledge in different disciplines.

[1] In discussing Web distributions we will use the terminology recommended by Vaughan (2005, p. 949) for types of hyperlinks. According to this terminology, an "inlink,"—a.k.a., "backlink"—is a link coming into a Web site, document, or page. In citation terms it is the equivalent of a "citation." On the other hand, an "outlink"—a.k.a., "outgoing link"—is a link going from the Web site, document, or page. In citation terms it is the equivalent of a "reference." Then there are "external links" (links coming from outside the item being linked) and "internal links" (links coming from within the item being linked). The latter can be thought of as "self-citations."

**The Yule-Simon distribution**

The Yule-Simon distribution had its origins in an effort by British biometricians to model Darwinian evolution mathematically. It was initially developed in the 1920s by Yule, in collaboration with botanist John Christopher Willis, whose theories on the distribution of species by genera and area Yule mathematically modeled. Willis believed that evolution proceeded by saltation or mutations (Turrill, 1958, pp. 354-355), making the appearance of a new species a Poisson event. Willis and Yule (1922) jointly previewed their distributional theories in an article in *Nature,* in which they expressed their two main guiding principles. The first was "Age and Area," by which Willis and Yule postulated that the area or space occupied in a country or the world by the species of a genus was proportional to their respective ages. The second was the twin principle of "Size and Space," by which Willis and Yule stipulated that the area or space occupied by related genera correlated with the number of species comprising the genera. Putting these two principles together, Willis and Yule (1922) stated that age, area (or space), and size go together and that, since age is the only operative factor of the three, whatever phenomena are shown by size should be similar to those shown by space (p. 177).

In the terminology of Willis, genera are posited to be distributed by both size (number of component species) and area in “hollow curves.” In actuality, these “hollow curves” are negative exponential J curves. This was demonstrated by Willis and Yule (1922), who found that the form of the frequency distribution for sizes of genera followed the rule that the logarithm of the number of genera plotted to the logarithm of the number of species yields a straight line with a negative slope. Figures 1 and 2 below are replications of their graphs of these “hollow curves” and the result of their logarithmic transformation.

To explain Willis’ evolutionary theory in terms of mathematical statistics, Yule (1925) developed a Poisson model in which evolutionary saltations or mutations were treated as random events occurring over time. An important aspect of this model—perhaps the most important—is its ability to show cumulative effects over time. According to Yule’s model, the number of species (i.e., size of a genus) increases in a geometric progression: if there is 1 species at time zero, there will be 2 species at time one, 4 species at time two, 8 species at time three, 16 species at time four, etc. This made “the doubling-period for species within the genus” (p. 25) the natural unit of time. Yule (1925) provided a generic demonstration of his model functioning over time with a hypothetical set of 1,000 genera operating over 6.28 doubling-periods (pp. 43-50). During this period, the range of the genera by size increased from a maximum of two genera with more than nine species to a maximum of 116 genera with more than 50 species, whereas the number of “monospecific” genera decreased from 571 to 343. As this process developed, the “hollow curve” increasingly approximated a negative exponential J curve. Yule demonstrated this with a series of graphs, where the distribution observed at each doubling-period was plotted on the

**Figure 1. Examples of distributional "hollow curves" posited by Willis in his evolutionary theory.**

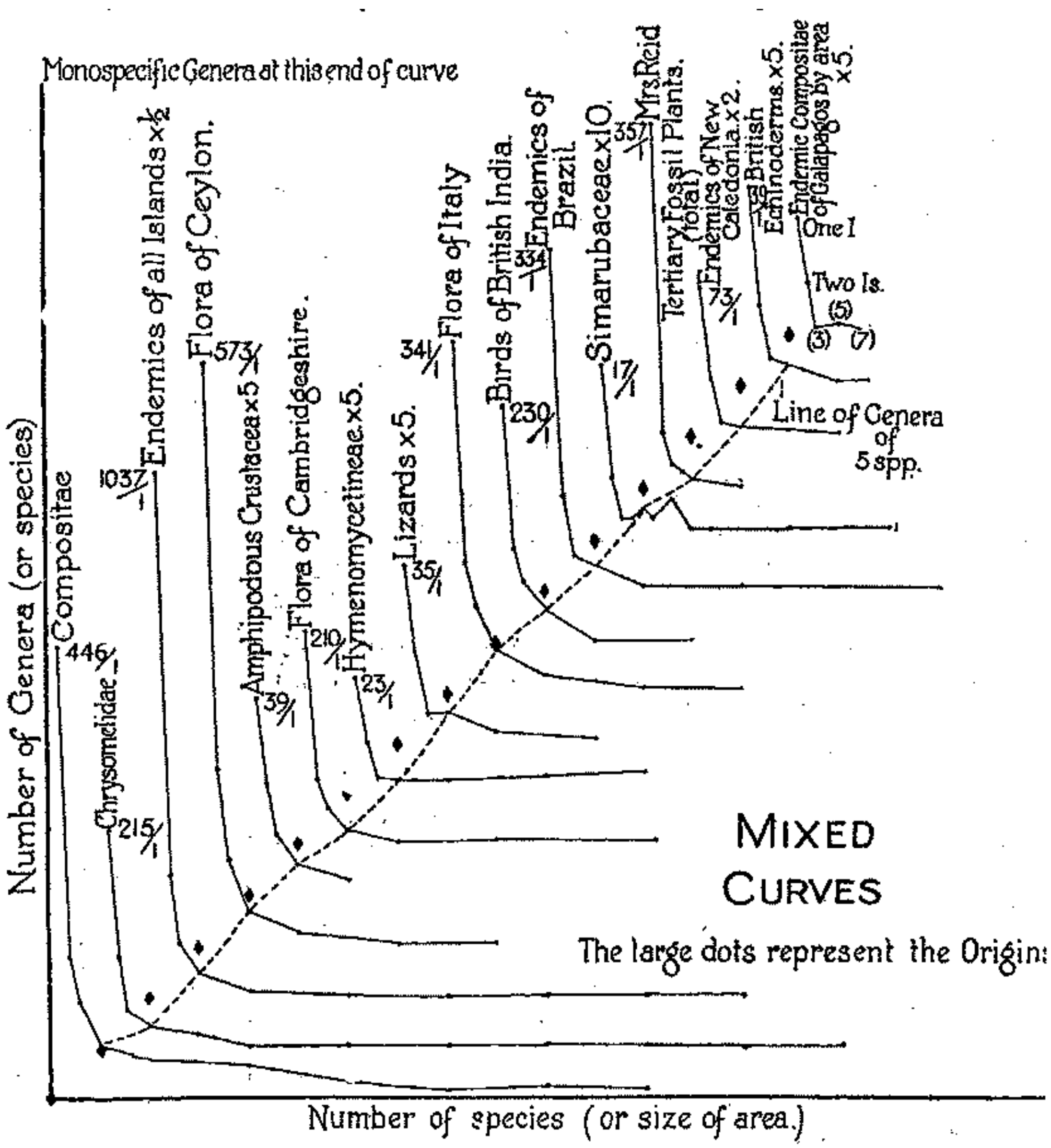

Note. From "Some statistics of evolution and geographical distribution in plants and animals, and their significance," by J. C. Willis and G. U. Yule, 1922. *Nature*, 109 (2728), p. 177.

**Figure 2. Logarithmic plot of a Willis "hollow curve" of number of genera to number of species**

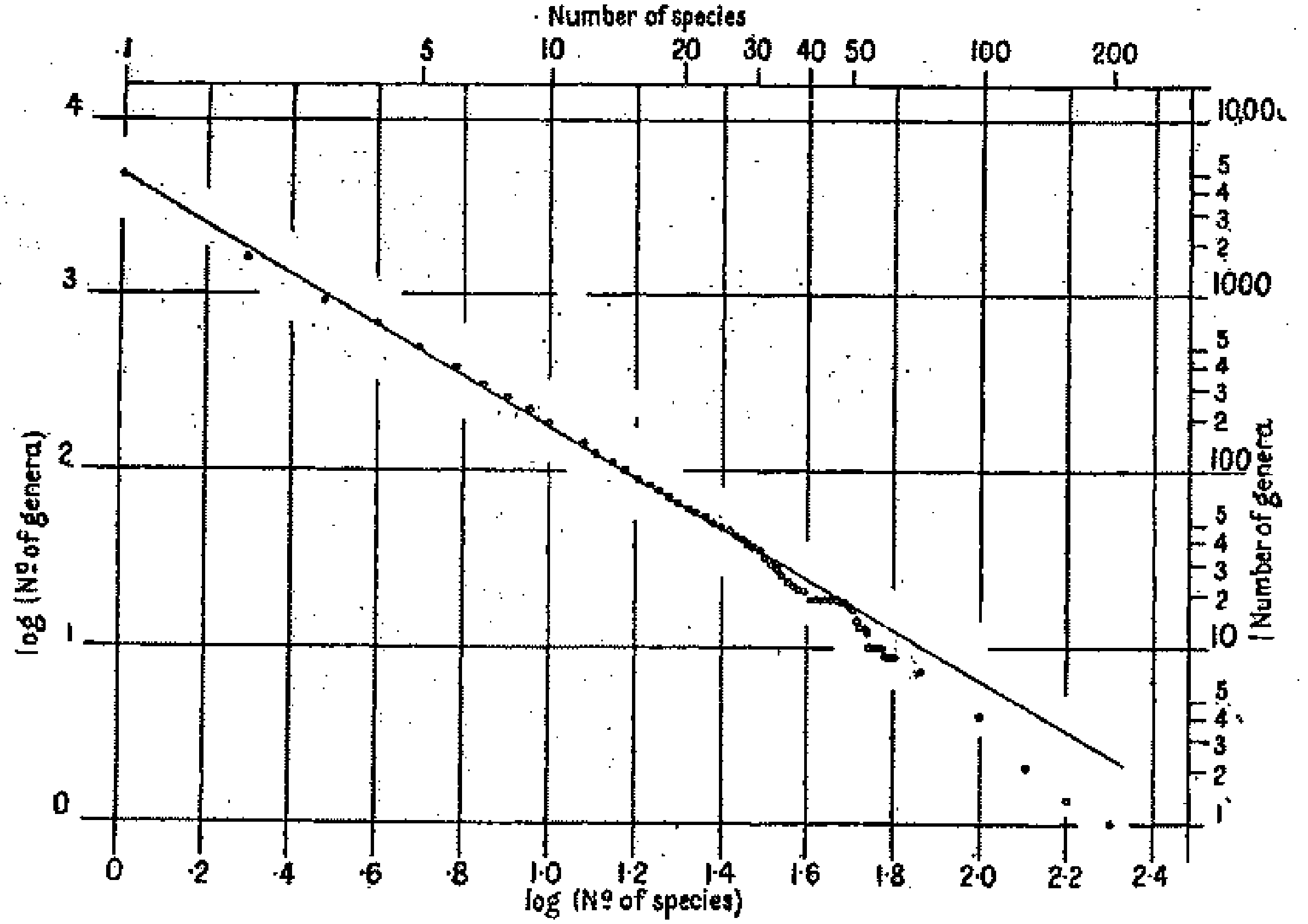

FIG. 2.—Log. curve for all flowering plants.

Note. From "Some statistics of evolution and geographical distribution in plants and animals, and their significance," by J. C. Willis and G. U. Yule, 1922. *Nature*, 109 (2728), p. 178.

**Figure 3. Yule's demonstration of his Poisson model of Willis' distributional theory of evolution with 1,000 hypothetical genera over 6.28 doubling-periods.**

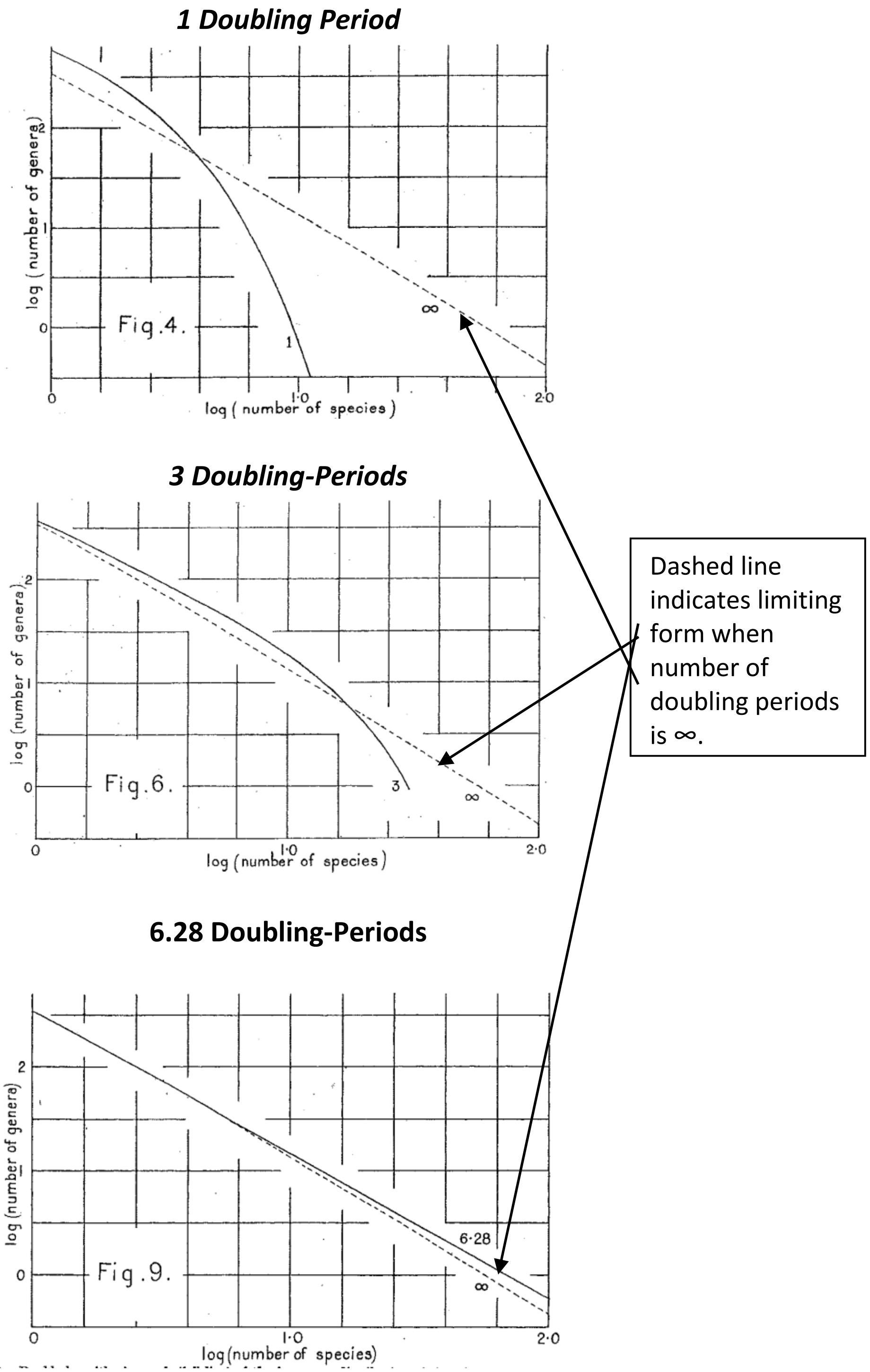

Note. From "A mathematical theory of evolution, based on the conclusions of Dr. J. C. Willis," by G. U. Yule, 1925. *Philosophical Transactions of the Royal Society of London*, Series B, *Containing Papers of a Biological Character*, 213, pp. 47-49.

logarithmic scale against a dashed line representing the limiting form when the number of doubling-periods equaled infinity. Yule's graphs for 1, 3, and 6.28 doubling-periods are shown in Figure 3 above. Here it can be seen that over time the "hollow curve" increasingly approximated the limiting negative exponential J curve represented by the straight dashed line with a negative slope. Needless to say, the positive skew of the observed distribution was geometrically increasing as the observed curve approximated the limiting curve.

Yule's model played no role in evolutionary theory. Its importance was not realized until it came to the attention of Herbert Simon (1955), winner of the 1978 Nobel Prize in economics, who gave it the name "the Yule distribution." Simon (1957) states that he was led to the Yule distribution by the work of Harvard philologist George Zipf, who reported remarkable regularities in such phenomena as the distribution of cities by size, the relation of rank order to frequency of word occurrences, the distribution of the frequencies of scientists' publications, etc. (p. 97). Zipf (1949) tried to bring these under a single mathematical relationship, devising a theoretical explanation he called "The Principle of Least Effort." Zipf specified that, according to this principle, in solving his immediate and estimated future problems a person "will strive to minimize the *probable average rate of his work-expenditure* (over time)" (p. 1). Simon (1957) described Zipf's principle as "a pseudo-explanation" and his data as "more irritating than a grain of sand in an oyster," because "the relations were undoubtedly genuine, and the purported explanation undoubtedly spurious" (p. 97).

Until then, Simon had worked primarily with deterministic models, but consideration of Zipf's problem convinced him that Zipf's regularities were statistical. The result was Simon's seminal paper on skew distribution functions. In this paper, Simon (1955) states that his purpose is to analyze a class of distributions that appears in a wide range of empirical data, particularly data describing sociological, biological, and economic data. According to him, these distributions appear so frequently in phenomena so diverse that one is led to the conjecture that the only property they can have in common is a similarity in the structure of the underlying probability mechanisms. He describes these curves as "J-shaped, or at least highly skewed, with very long upper tails" (p. 425), i.e., negative exponential curves. He then introduces what he terms "the Yule distribution" (p. 426), based upon the incomplete beta function. In a book on skew distributions and the size of business firms, Ijiri and Simon (1977) give the following succinct, nonmathematical description of the stochastic process underlying the Yule distribution:

> This process starts with a few elements all of unit size in order to initialize the population. At each epoch, $\tau$, the aggregate size of the population (the sum of the sizes of the elements of the population) is increased by one unit. Two assumptions govern the rule on which element is to receive the unit increment. The first assumption states that there is a constant probability that the unit goes to a new element not previously in the population. If this happens, a new element of unit size is created in the population. If the unit does not go to a new element, it goes to an old element. In this case, the selection of the old element to which the unit is given is governed by the second assumption: the probability that the selected element is of size $i$ (before the unit is added) is proportional to the aggregate size of all elements with size $i$ ($i$ times the number of elements with size $i$). The second assumption incorporates Gibrat's law of proportionality. Hence, big or small, each element has the same chance of growing at any given percentage in a given period. (p. 66)

Ijiri and Simon (1977) define the Gibrat law of proportionality as "the assumption that the expected percentage rate at which something will grow is independent of the size it has already attained," stating that "the Gibrat assumption virtually guarantees that the distribution will be highly skewed, with a long upper tail" (p. 24). Here it should be stated that the probability of a new unit going to an element of a given size *i* is proportional to the aggregate size of the *i* which stipulates that the Yule distribution is based upon a success-breeds-success mechanism, because the bigger and more numerous the elements of a given size *i,* the higher the probability of a new unit going to one of them. Simon (1955) showed that the Yule model provided good fits to empirical data sets not only of biological species but also city sizes, income distribution, word frequencies, and—most interestingly—papers authored by chemists, economists, mathematicians, and physicists. This paper is of historical importance for fitting the Yule-Simon model to empirical distributions of words by frequency and publications by scientists. It marked the first direct connection of a stochastic model to informetrics and its laws.

The Yule-Simon distribution is of great potential importance for both informetrics and webometrics. In respect to the first, informetrics is distinguished by a number of empirical laws, some of which have precise mathematical formulations. To the latter category belong Lotka's Law of Scientific Productivity, Bradford's Law of Scattering, and Zipf's Law of Word Frequency. In her authoritative review of these laws, Wilson (1999) states that the commonality shared by these laws is that "the form of the distribution of numbers of items over the number of sources, when appropriately displayed, is variously described as extremely right-skewed, reversed-J-shaped, or quasi-hyperbolic" (p. 165). According to her, one way to deal with informetric regularities has been through various forms of the inverse power law.

There have been a number of investigations of the relationship of the Yule-Simon model to informetric laws. Chen and Leimkuhler (1986) derived a common functional relationship among Lotka's Law, Bradford's Law, and Zipf's Law, demonstrating that these three laws are equivalent. Chen (1989) then showed that Lotka's Law can be derived from the generating mechanism of the Yule-Simon model. Chen, Chong, and Tong (1994) used a simulation algorithm based on the Simon-Yule model to conduct computational experimentation on the three informetric laws, finding that the probability of a new entry, be it constant or decreasing, determines the characteristics of all three distributions. In her review of the literature on the Yule-Simon approach to informetrics, Wilson (1999) particularly noted its ability to monitor the evolution of the distribution over time, and stated that the value of this approach seems assured (pp. 195-197).

In respect to webometrics, the Yule-Simon distribution also appears to be a feasible model for the type of distribution dominating the Web. This is evident in the seminal book *The Laws of the Web: Patterns in the Ecology of Information* by Huberman (2001). According to Huberman, Web distributions are governed by a power law; for example, the probability of finding a Web site with a given number of pages (n) is proportional to $1/n^\beta$, where $\beta$ is a number greater than or equal to one. These distributions are scale free, so that the number of pages per site or number of links per site being distributed according to a power law is a universal feature of the Web, holding everywhere no matter what the scale or type of site. Seeking an analogy for the power-law structure of the Web, Huberman (2001) referred to Zipf's Law, which is specifically modeled by the Yule-Simon distribution (pp. 30-31).

**Publication impact of Nobelists in Chemistry**

Data on the laureates and medalists were downloaded in September, 2011. The data were loaded into Excel spreadsheets with the aid of the Publish or Perish (PoP) computer program developed by Anne-Wil Harzing (described in Harzing, 2010). This program is freely available at http://Web.harzing.com.[2] For purposes of simplification, the analysis of this chapter will be restricted to publications comprised of the researchers' *h*-indexes or the upper ranks of Google Scholar inlinks, which are automatically calculated by PoP. One reason for this is that the Google search engine, like others, delivers its most relevant sets at the upper ranges.

The *h*-index has become one of the leading measures for the evaluation of scientists. It was created by J. E. Hirsch, a physicist at the University of California at San Diego. In his initial formulation, Hirsch (2005) defined his *h*-index thus: "A scientist has index *h* if *h* of his or her *Np* papers have at least *h* citations each and the other (*Np* - *h*) papers have ≤ h citations each" (p. 16569), where *Np* is the number of papers published over *n* years. Hirsch (2007) then modified his initial formulation in the following important way: "The *h* index of a researcher is the number of papers **coauthored** [emphasis added] by the researcher with at least *h* citations each" (p. 19193). The amendment to co-authorship is of extreme importance for a field such as chemistry, which is highly collaborative. GS retrieves works attributable to a scientist no matter what the position of this scientist in the authorship order of the publication (be it primary author, secondary author, or even editor, as in the case of books). For the most part, what is actually being measured in chemistry is not the inlinks to the publications authored by the Nobelists, but rather links to publications produced by the author collective in which the Nobelist functioned.

Publication and inlink data were retrieved from GS for five chemistry Nobelists (listed in Table 1 below). A major factor in their selection was the year of their prize, allowing the researchers to judge the possible effect of time. Thus, three of the Nobelists won their prizes a decade apart—Corey in 1990, Heeger in 2000, and Negishi in 2010—and three of the Nobelists

[2] This is a program that has arguably established statistical control over Google Scholar. The data on the chemists were downloaded by Harzing and provided to the primary author of this chapter for analysis.

| Table 1. Google Scholar inlinks to the works of the winners of the Nobel prize in distribution truncated at the h-index. | | | | | | |
|---|---|---|---|---|---|---|
| **Prize Winner** | | **H-Index** | **Inlink Range** | | | **Total Inlinks** |
| **Name** | **Year** | | **Minimum** | **Maximum** | **Total Range** | |
| Elias J. Corey | 1990 | 97 | 97 | 1225 | 1129 | 23216 |
| Alan J. Heeger | 2000 | 123 | 123 | 3321 | 3199 | 49266 |
| Osamu Shimomura | 2008 | 50 | 52 | 929 | 878 | 5852 |
| Ada E. Yonath | 2009 | 36 | 36 | 637 | 602 | 4381 |
| Ei-ichi Negishi | 2010 | 49 | 50 | 797 | 748 | 6692 |

won their prizes only one year apart—Shimomura in 2008, Yonath in 2009, and Negishi in 2010—the three most recent years at the end of the 30-year period. Of great significance is the fact that the five chemists won their prizes at an advanced age, when they were reaching the end of their careers. Corey was the youngest at the age of 62, whereas Shimomura was the oldest, at 80. The average age of the chemists at the time of their prize was 70.

Burrell (2007) has proposed a statistical model for the *h*-index, according to which, all other things being equal, Hirsch's *h*-index is: (i) approximately proportional to current career length; (ii) approximately a linear function of the logarithm of the author's productivity rate; and (iii) approximately a linear function of the logarithm of the mean citation rate for the author, for moderate citation rates. The findings on the Nobelists provide empirical evidence of the correctness of Burrell's model. Thus, the *h*-indexes of the five Nobelists were as follows: Corey—97; Heeger—123; Shimomura—50; Yonath--36; and Negishi—49. There appears to be a time effect, because Corey and Heeger, who won their prizes in 1990 and 2000 respectively, have *h*-indexes of 97 and 123 (respectively), whereas Shimomura, Yonath, and Negishi, who won their prizes from 2008 to 2010, have *h*-indexes ranging from 36 to 50. Of great significance in this respect is the greater inlink range of the Nobelists who won their prizes earlier. Thus, whereas the top works of Corey and Heeger have 1,225 and 3,321 inlinks respectively, the top works of Shimomura, Yonath, and Negishi respectively have 929, 637, and 797 inlinks. There is also a tendency for the total number of inlinks within the *h*-index range to increase over time. Thus, the total inlinks for Shimomura (2008), Yonath (2009), and Negishi (2010) ranged from 4381 to 6692, whereas the total inlinks for Corey (1990) and Heeger (2000) were 23,216 and 49,266, respectively.

Restriction of the analysis to the publications comprising the chemistry laureates' *h*-indexes truncated these distributions on the left. The shapes of these *h*-index-truncated distributions were structured and graphically explored with the aid of Excel. First, the truncated inlink range was divided by 25 to create inlink categories ("bins," per Excel terminology) that each comprised 4% of the truncated inlink range. Second, the Excel FREQUENCY function was utilized to allocate the works in accordance with the number of inlinks into their proper inlink category or bin. Finally, Excel Chart Tools were employed to create bar charts with the number of works on the y-axis and the inlink categories on the x-axis. The first charts in Figures 4 and 5 below show the results for Negishi (2010), the latest laureate, and Corey (1990), the earliest laureate, revealing the possible effect of time. Their charts are representative of what was found for the three other intermediary Nobelists. There are a number of important things that should be noted about these distributions. First, all of the distributions are highly and positively skewed with a long tail (or asymptote) to the right. There is a concentration of publications in the lowest 4% of the truncated inlink range, indicating that the publications located in the exponentially increasing asymptote to the right are the laureates' most significant ones. Second, comparison of these distributions with those in Figure 1 reveals their close resemblance to the "hollow curves" which were posited by Willis in his evolutionary theory and are modeled by the Yule-Simon distribution.

**Figure 4.** ***h*-index truncated distribution of publications retrieved for Ei-ichi Negishi. 2010 chemistry Nobelist, from Google Scholar (GS) across GS inlink categories and linear fit to it.**

**Truncated distribution**

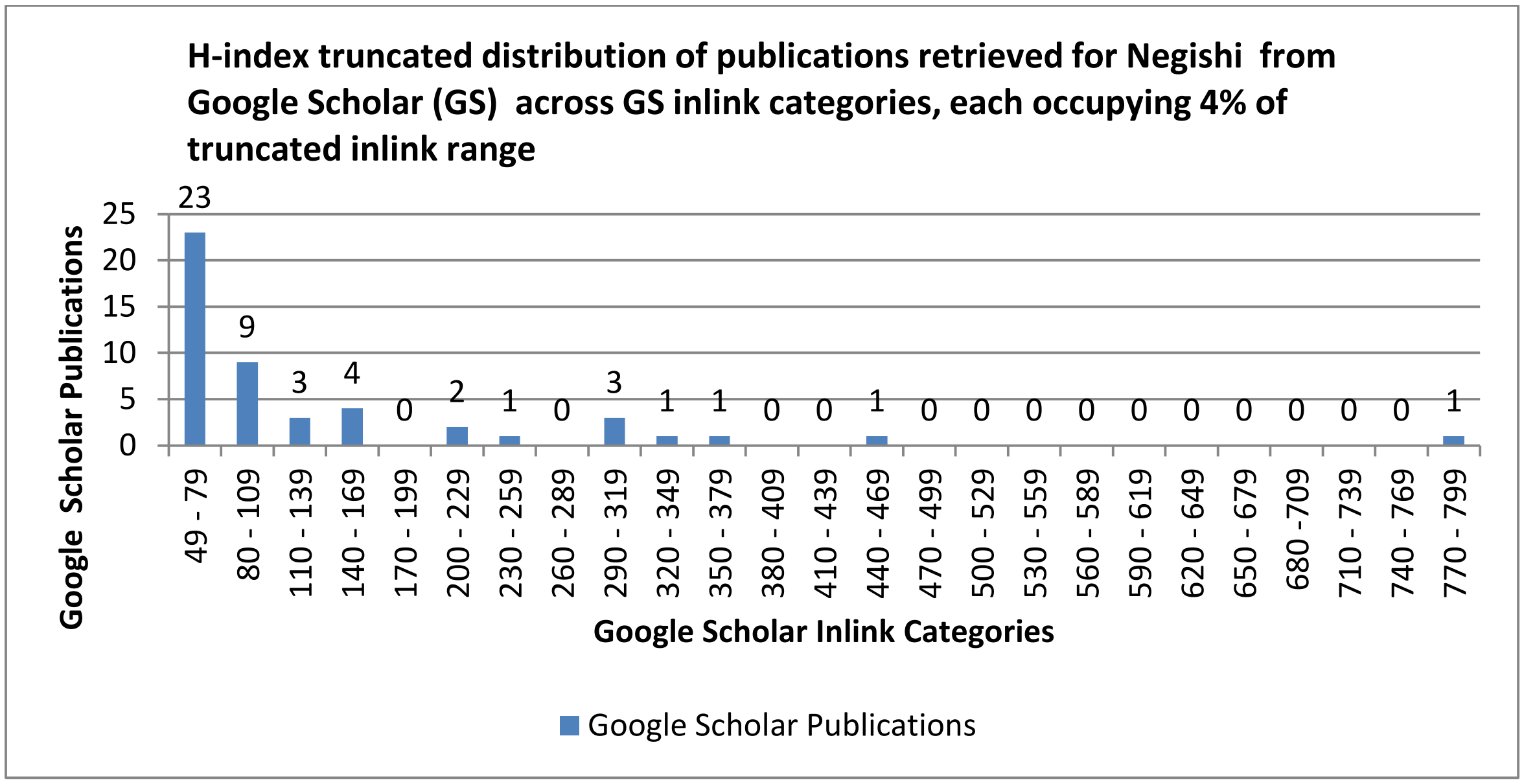

**Linear fit to truncated distribution**

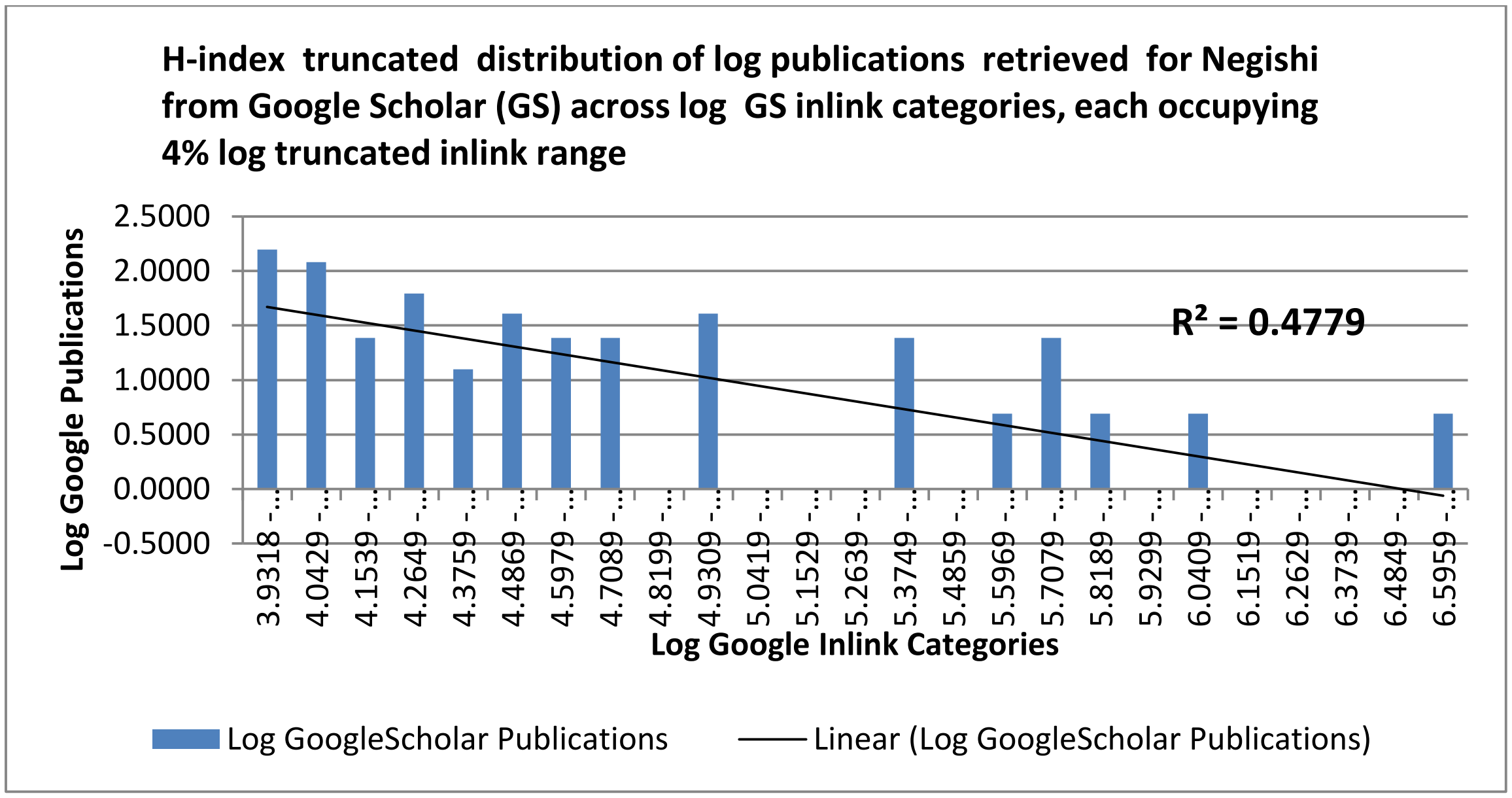

**F = 21.05. Fit is significant at the 1% level**

**Figure 5.** ***h*-index truncated distribution of publications retrieved for Elias K. Corey, 1990 chemistry Nobelist, from Google Scholar (GS) across GS inlink categories and linear fit to it.**

**Truncated distribution**

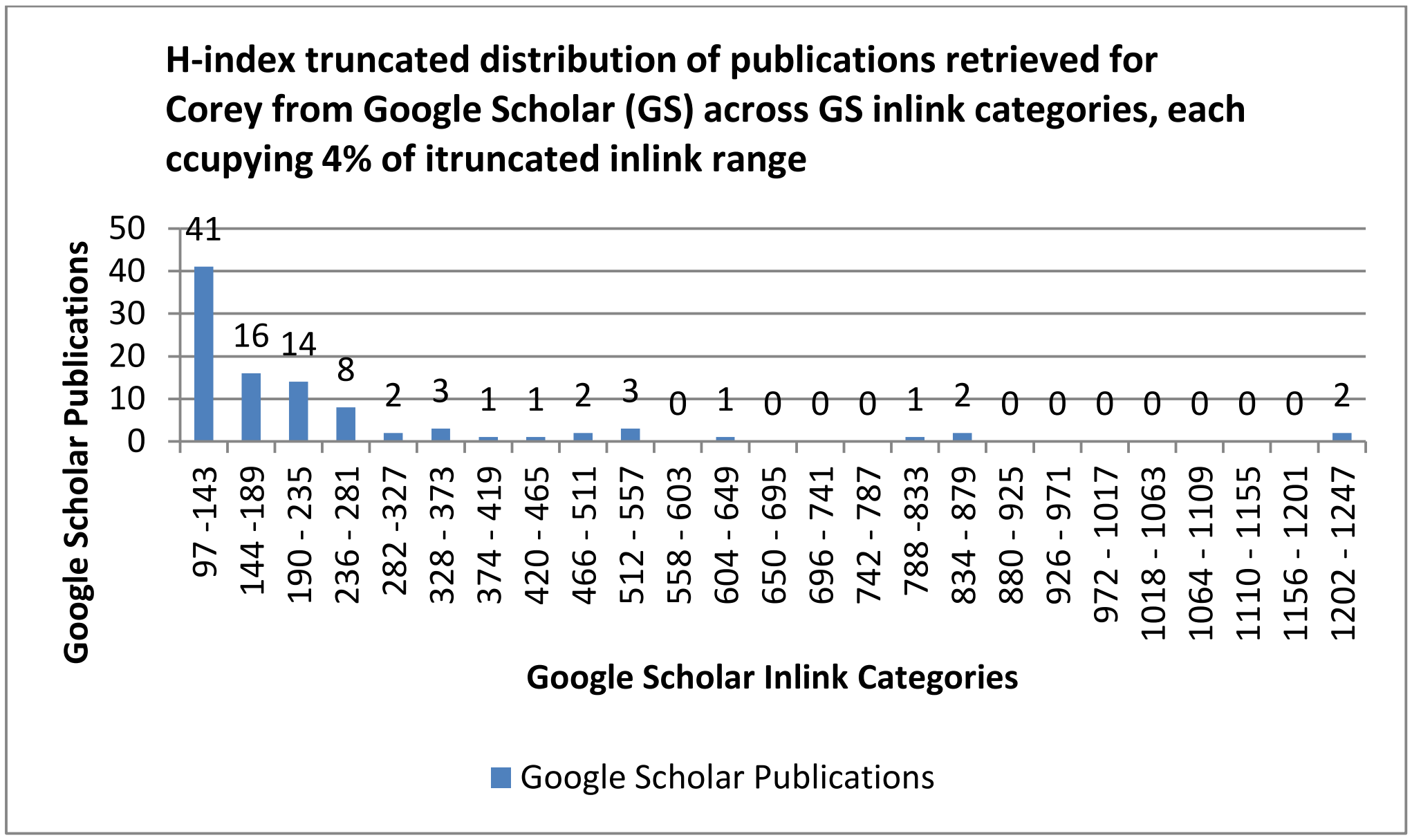

**Linear fit to truncated distribution**

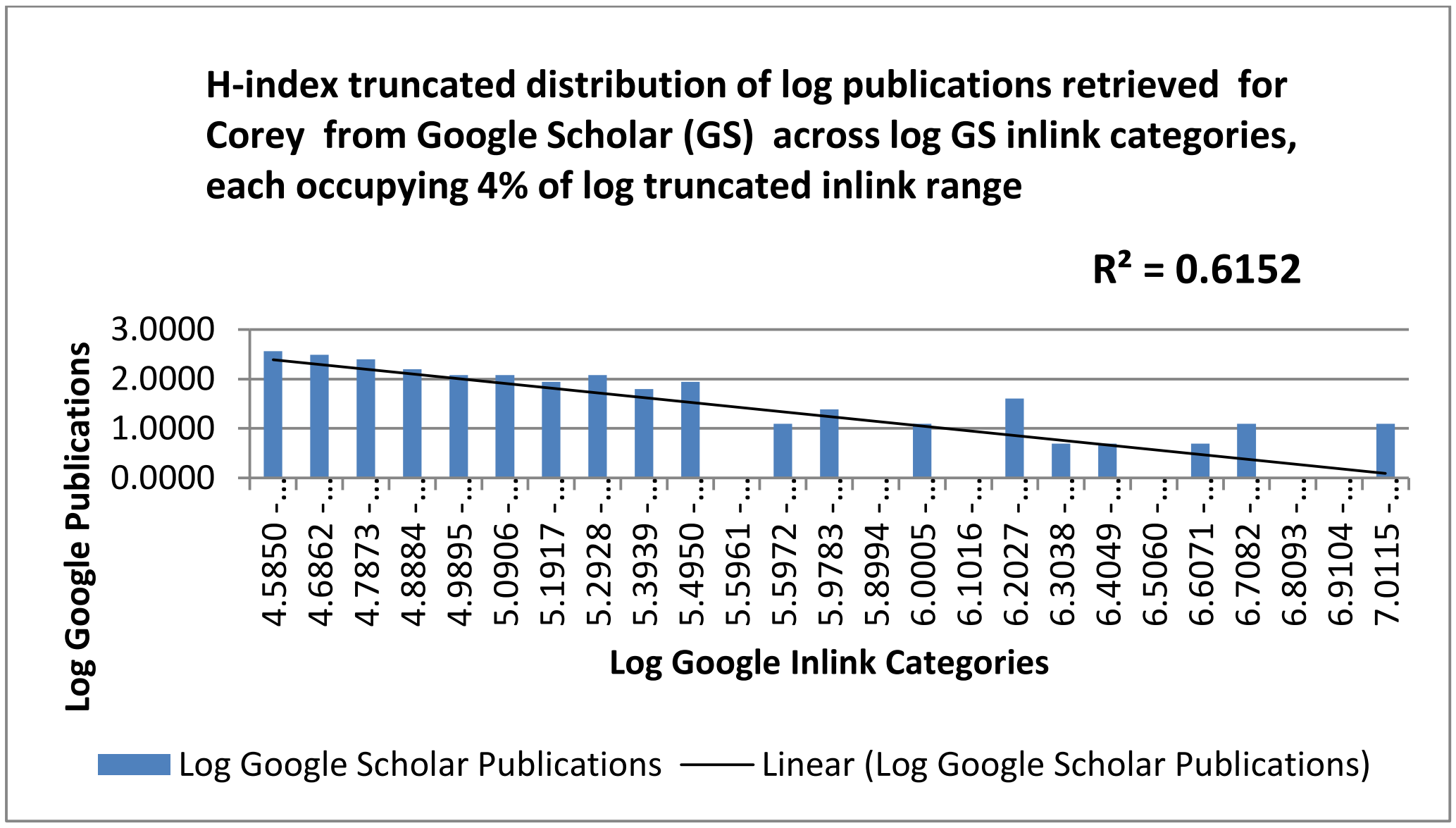

**F = 36.77. Fit is significant at the 1% level**

On the basis of the graphical evidence, the Yule-Simon distribution appears feasible not only as a model of the productivity of scientists in number of publications, where it was first applied to an informetric distribution governing scientometrics, but also for the distribution of citations or inlinks over the publications of a given scientist. A key feature of this model was demonstrated by Yule himself in a series of graphs, of which key ones are replicated in Figure 3. The distribution develops over time in a series of doubling periods, going to a limit where it can be said to be complete. If, for example, both the species and their size or area are logarithmically transformed, the progress of this development can be measured by calculating the fit of the data to a hypothetical straight line with a negative slope. For this reason, it seems theoretically possible to measure how well the works of a given scientist have been incorporated into the body of knowledge of her/his field. This concept was tested on the *h*-index-truncated distributions of the inlinks over the publications of the five chemistry Nobelists retrieved by Google Scholar, using the Excel function LINEST. To conduct the tests, the data were structured in the following way. First, the numbers of both the publications and the inlinks were transformed by taking their natural log. In cases where there were zero counts, the transformation was the natural log plus one (the log of zero is nonsensical, whereas the log of one is zero, resulting in a ratio scale with zero as its origin). Second, the logarithmic inlink range was divided by 25 to create inlink categories (or bins), each comprising 4% of the logarithmic inlink range. Once again, the Excel FREQUENCY function was utilized to allocate the works in accordance with their logged inlinks into their proper logged inlink category (or bin). Third, LINEST does regression analysis, where it is necessary to distinguish between the dependent variable Y and the independent variable X. The independent variable X was calculated by transforming the number of works in a given category or bin by taking the natural logarithm plus one, whereas the dependent variable Y was created by calculating the midpoint of each of the 25 logarithmic inlink categories. Fourth, Excel Chart Tools were used to create bar graphs on which the independent variable—log number of works in a given bin—is on the vertical y-axis, whereas the horizontal x-axis demarcates the logarithmic inlink categories; the midpoint of the x-axis estimates the dependent variable or inlinks to the works in the bin.

The Excel function LINEST regresses the independent variable on the dependent variable and makes a number of important calculations, the results of which are summarized in Table 2 below and shown in Figures 4 and 5. LINEST uses the "least squares" method to calculate the linear trendline that best fits the data and the coefficient of determination or R^2 that measures the goodness of this fit. An R^2 equaling zero indicates no fit at all, whereas an R^2 equaling one indicates a perfect fit. The closer R^2 is to one, the better the fit. LINEST also calculates the F statistic and a degrees of freedom (df) statistic, with which it is possible to determine the significance of the fit. Due to the way the data were structured, the number of data points (or n) always equaled 25, and df always equaled 23. In order to enter published tables of the F distribution to determine the level of significance, one needs to know the v1 and v2 degrees of freedom. LINEST gives the following formulae for these:

$$v1 = n - df - 1 = 25 - 23 - 1 = 1$$

$$v2 = df = 23$$

At this point, an F at or above 4.28 is significant at the 0.05 level, and an F at or above 7.88 is significant at the 0.01 level. Thus, the higher the F, the more significant the fit. Using the example of Negishi, inspection of Table 2 reveals that his R^2 equals 0.4779 with an F equal to 21.05, which is significant at the 0.01 level.

| Table 2. Linear fits (R^2) of log Google Scholar (GS) publications of Nobel prize winners in chemistry to their log GS inlinks in the distribution truncated at the h-index. | | | |
|---|---|---|---|
| **Prize Winner** | **Year** | **R^2*** | **F**** |
| Elias J. Corey | 1990 | 0.6152 | 36.77 |
| Alan J. Heeger | 2000 | 0.7397 | 65.35 |
| Osamu Shimomura | 2008 | 0.6501 | 42.73 |
| Ada E. Yonath | 2009 | 0.4685 | 20.28 |
| Ei-ichi Negishi | 2010 | 0.4779 | 21.05 |

* The Excel LINEST function uses regression analysis to calculate the linear trendline, using the "least squares" method to find the straight line that best fits the data. R^2--or the coefficient of determination--runs from 0 or no fit of the line to the data to 1 or a perfect fit of the line to the data.

** For significance of the fit, LINEST not only produces the F statistic but also a degrees of freedom (df) statistic that enables calculating the v1 and v2 degrees of freedom necessary to enter published F-distribution tables at the proper place to determine the significance level. The formulas are v1 = n-df-1 and v2 = df, where n is the number of data points and df is the LINEST degrees of freedom. Our data was organized in such a way that n always equaled 25 and df always equaled 23. Therefore, v1 always equaled 1, and v2 always equaled 23. As a result, the same level of significance is applicable to all the tests above. To be significant, F has to 4.28 and above at the 5% level or 7.88 and above at the 1% level.

An inspection of Table 2 reveals that both the R^2 and its F significance appear to increase with time. For the first three Nobelists—Corey (1990), Heeger, (2000), and Shimomura (2008)—R^2 ranges from 0.6152 to 0.7397, and the F statistic ranges from 36.77 to 65.35. In contrast, for the last two Nobelists—Yonath (2009) and Negishi (2010)—the R^2s with their F significances are respectively 0.4685 at 20.28 and 0.4779 at 21.05. In a study of citations, paradigm shifts, and Nobel prizes, Mazloumian, Eom, Helbing, Lozano, and Fortunato (2011) observed that winning the Nobel Prize not only increases the number of citations for "landmark papers" but also tends to "boost the citation rates of their previous publications" (p.1). It seems that the winning of the prize itself may also have this effect, with a lag period of about 2 years.

This is demonstrated by Yonath (2009) and Negishi (2010) having much lower fits than Shimomura (2008), who more resembles Corey (1990) and Heeger (2000) in this respect.

**Publication impact of the Fields medalists in Mathematics**

The Fields medal differs from the Nobel prize in two important respects. First, unlike the Nobel prize, of which one, two, or three are awarded every year in each discipline, the Fields medal is awarded to two, three, or four mathematicians once every four years. This temporal spacing dictated the selection of the five Fields medalists listed in Table 3 below. The authors wished to include Fields medalists that were recent, as well as those who were well-established in their careers. Therefore, they chose the years 1986, 1998, and 2010 (equivalent to a period of 12 years of Nobel awards). The medalists were bar-belled by selecting two (Donaldson and Faltings) for the earliest year (1986), one (Kontsevich) for the middle year (1998), and two (Chau and Lindenstrauss) for the most recent year (2010), increasing the possibility of temporal effects. The second major way the Fields medal differs from the Nobel prize is that, unlike the latter, the former cannot be awarded as the culmination of a career, but rather is granted to mathematicians of (at most) 40 years of age. Thus, of the five medalists listed in Table 3, Donaldson was youngest at the time of his medal at the age of 29, whereas Lindenstrauss was the oldest at the age of 40. The average of these five medalists was 35—half the average age of the five chemistry Nobelists.

As a discipline, mathematics is marked by a number of bibliometric characteristics that distinguish it from a natural science such as chemistry. Bensman, Smolinsky, and Pudovkin (2010) found a high degree of apparent randomness in the impact factor distribution of mathematics, and, in analyzing the bibliometric structure of the discipline, found it a disjointed discipline of isolated subfields with a weak central core of journals, reduced review function, and

| Table 3. Google Scholar inlinks to the works of the winners of the Fields Medal in distribution truncated at the h-index. | | | | | | |
|---|---|---|---|---|---|---|
| **Medalist** | | | **Inlink Range** | | | |
| **Name** | **Year** | **H-index** | **Minimum** | **Maximum** | **Total Range** | **Total Inlinks** |
| Simon K. Donaldson | 1986 | 33 | 36 | 1233 | 1198 | 6864 |
| Gerd Faltings | 1986 | 27 | 29 | 950 | 922 | 3694 |
| Maxim L. Kontsevich | 1998 | 31 | 37 | 1742 | 1706 | 8462 |
| Ngo Bao Chau | 2010 | 13 | 13 | 74 | 62 | 343 |
| Elon Lindenstrauss | 2010 | 15 | 17 | 150 | 136 | 759 |

long cited half-life. Whereas chemistry review articles by leading researchers play a major role in defining the paradigms of the discipline, this is not the case in mathematics, where the equivalent is the so-called "expository article," in which mathematicians in one subfield attempt to explain what is taking place in their field to mathematicians in other subfields. This situation was succinctly described by Singh (1997) in the following colorful fashion:

> Mathematics consists of islands of knowledge in a sea of ignorance. For example, there is the island occupied by geometers who study shape and form, and then there is the island of probability where mathematicians discuss risk and chance. There are dozens of such islands, each one with its own unique language, incomprehensible to the inhabitants of other islands. (p. 191)

The 2011 *Science Citation Index Journal Citation Reports* lists the subject category "Chemistry, Multidisciplinary" as having an aggregate cited half-life—or median age of the articles cited in that category that year—of 5.9 years and the subject category Mathematics as having one of greater than 10 years. To compound matters, Smolinsky and Lercher (2012) compared citation counts for award-winning mathematicians in different subfields, finding not only a pattern in which mathematicians working in some subfields had fewer citations than others who won the same award, but also that citation counts for different subfields did not match peer evaluation.

All of these factors are visible in the inlink data downloaded for the five Fields medalists from GS. Looking at Table 3, it is shown that medalists' *h*-indexes tend to increase over time, rising from a low of 13 to a high of 33 over the 24-year period. In this regard the mathematicians resemble the chemists, but the *h*-indexes of the latter were higher, ranging from a low of 36 to a high of 123 over the 30-year period. The pattern was basically the same for the maximum number of inlinks and the total number of inlinks. The indexes of both the mathematicians and the chemists had a tendency to rise over time, and those of the chemists tended to be higher than those of the mathematicians. The figures for the mathematicians were the following: lowest maximum – 74; highest maximum – 1,742; lowest total – 343; highest total – 8,462. The comparable figures for the chemists were: lowest maximum – 637; highest maximum – 3,321; lowest total – 4,381; highest total – 49,266.

The Fields medalists were graphed in the same fashion as the Nobel laureates so that their *h*-index-truncated distributions could be compared. The results are shown by the first charts in Figures 6 and 7 below. As with the chemists, one of the most recent medalists, Lindenstrauss (2010), was graphed together with one of the earliest, Faltings (1986). Whereas Faltings' graph resembles those of the chemists, the same cannot be said of Lindenstrauss' graph, which does not have the characteristics of a negative exponential curve with a long asymptote to the right. The differences between mathematics and chemistry as disciplines are made clear with the R^2 tests measuring the incorporation of the work of the prize winners into the knowledge corpus of their fields. These differences are manifest in Table 4 below. Here the R^2 of the two 2010 medalists are negligible and insignificant. Inspection of the charts in Figure 6 for Lindenstrauss—the first

**Figure 6. *h*-index truncated distribution of publications retrieved for Elon Lindenstrauss, 2010 Fields medalist from Google Scholar (GS) across GS inlink categories and linear fit to it.**

**Truncated distribution**

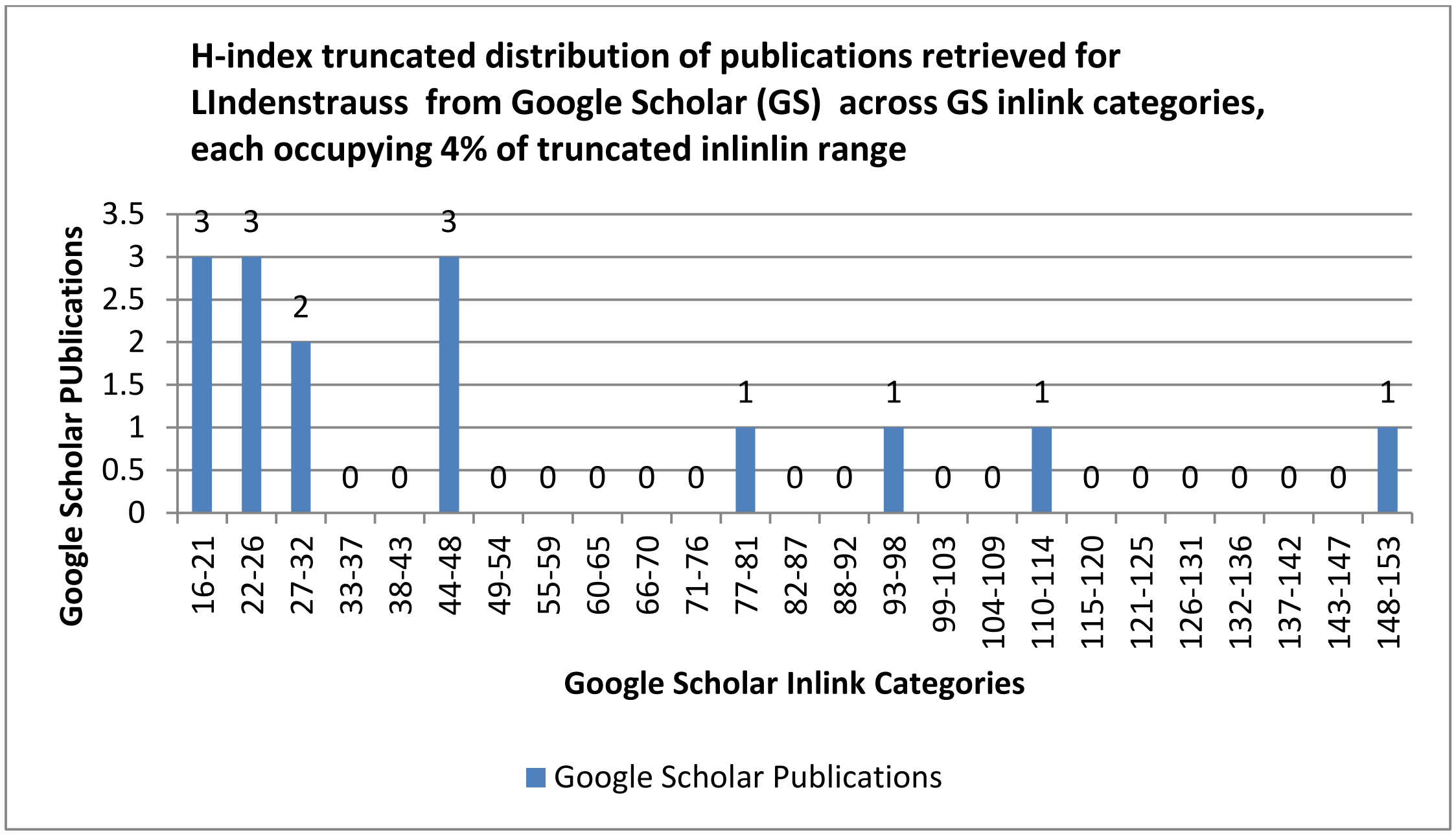

**Linear fit to truncated distribution**

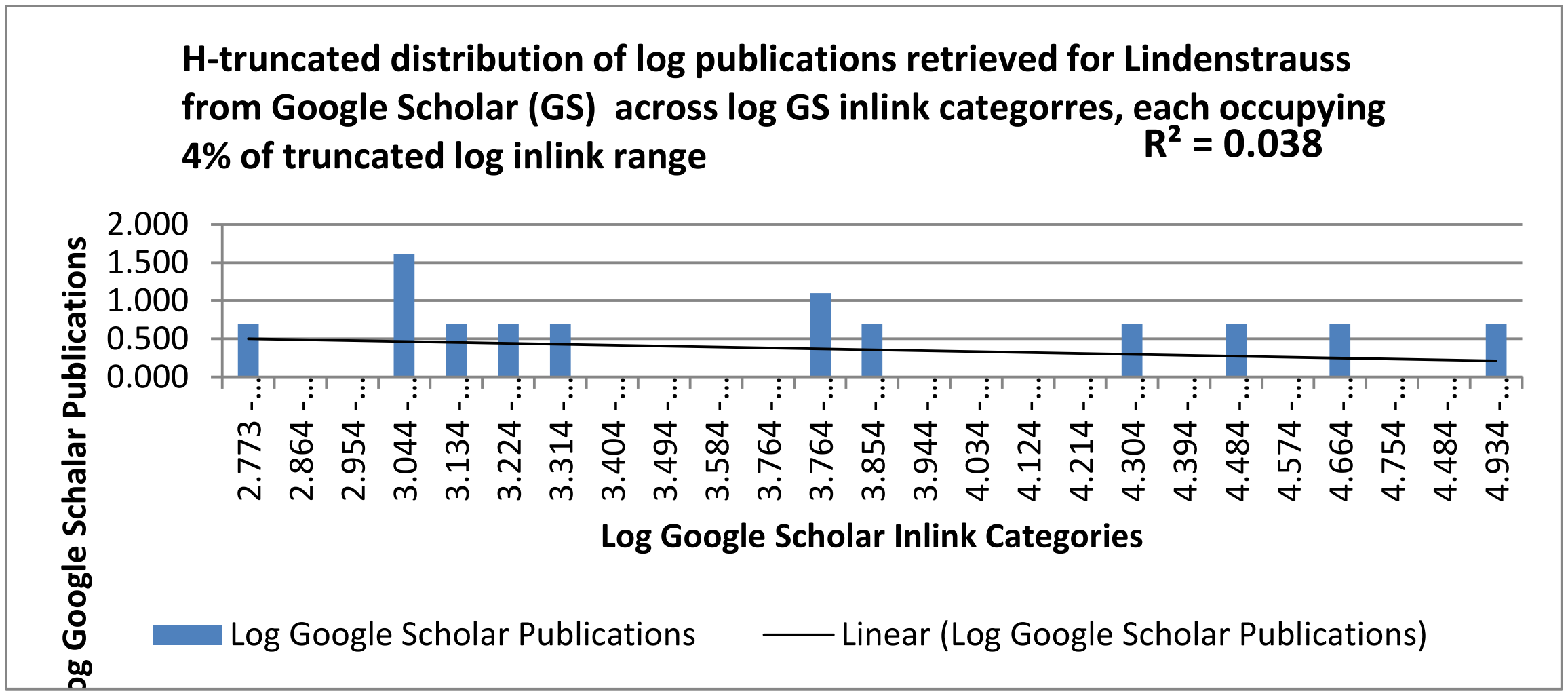

**F = 0.91. Fit is not significant.**

**Figure 7. *h*-index truncated distribution of publications retrieved for Gerd Faltings, 1986 Fields medalist from Google Scholar (GS) across GS inlink categories and linear fit to it.**

**Truncated distribution**

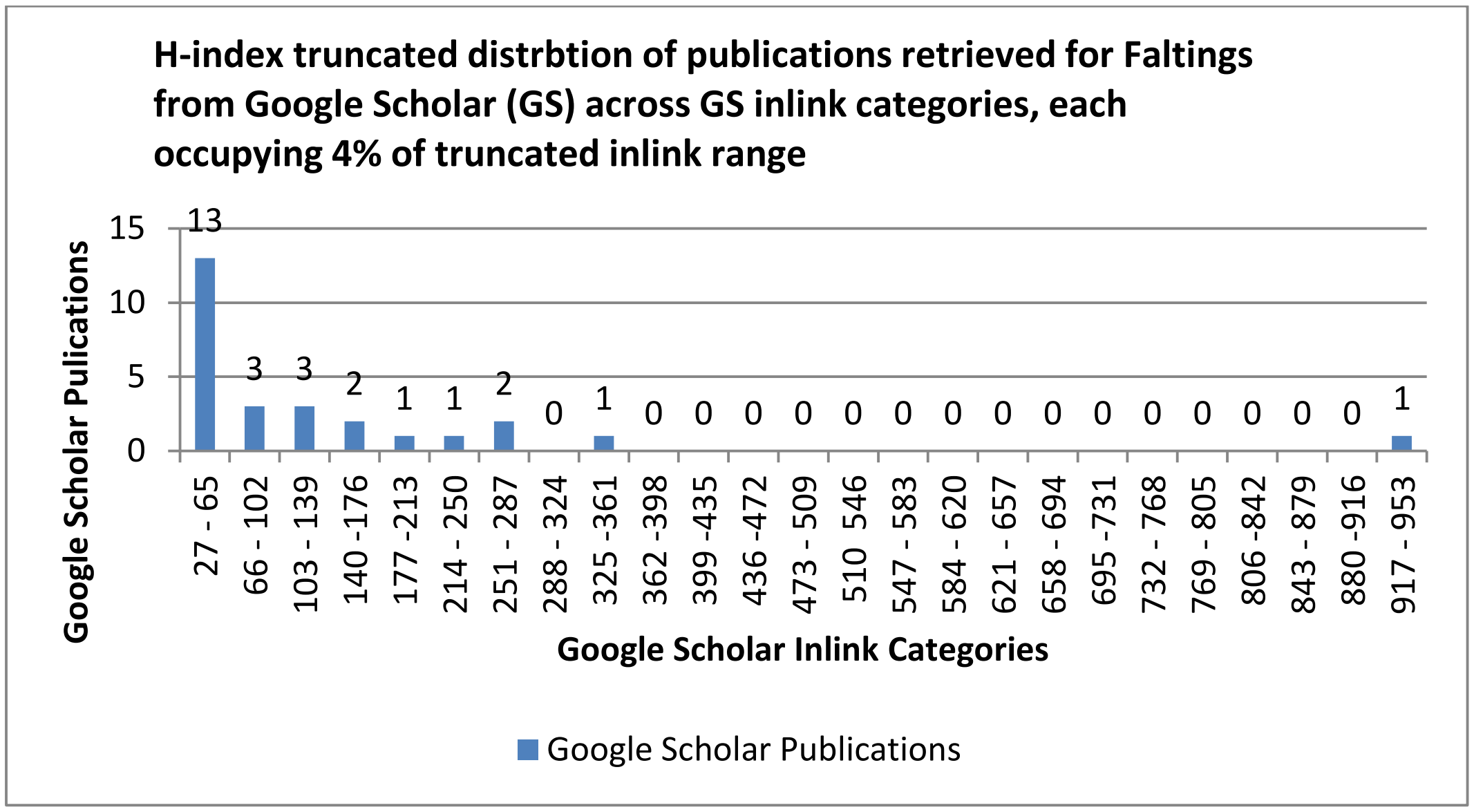

**Linear fit to truncated distribution**

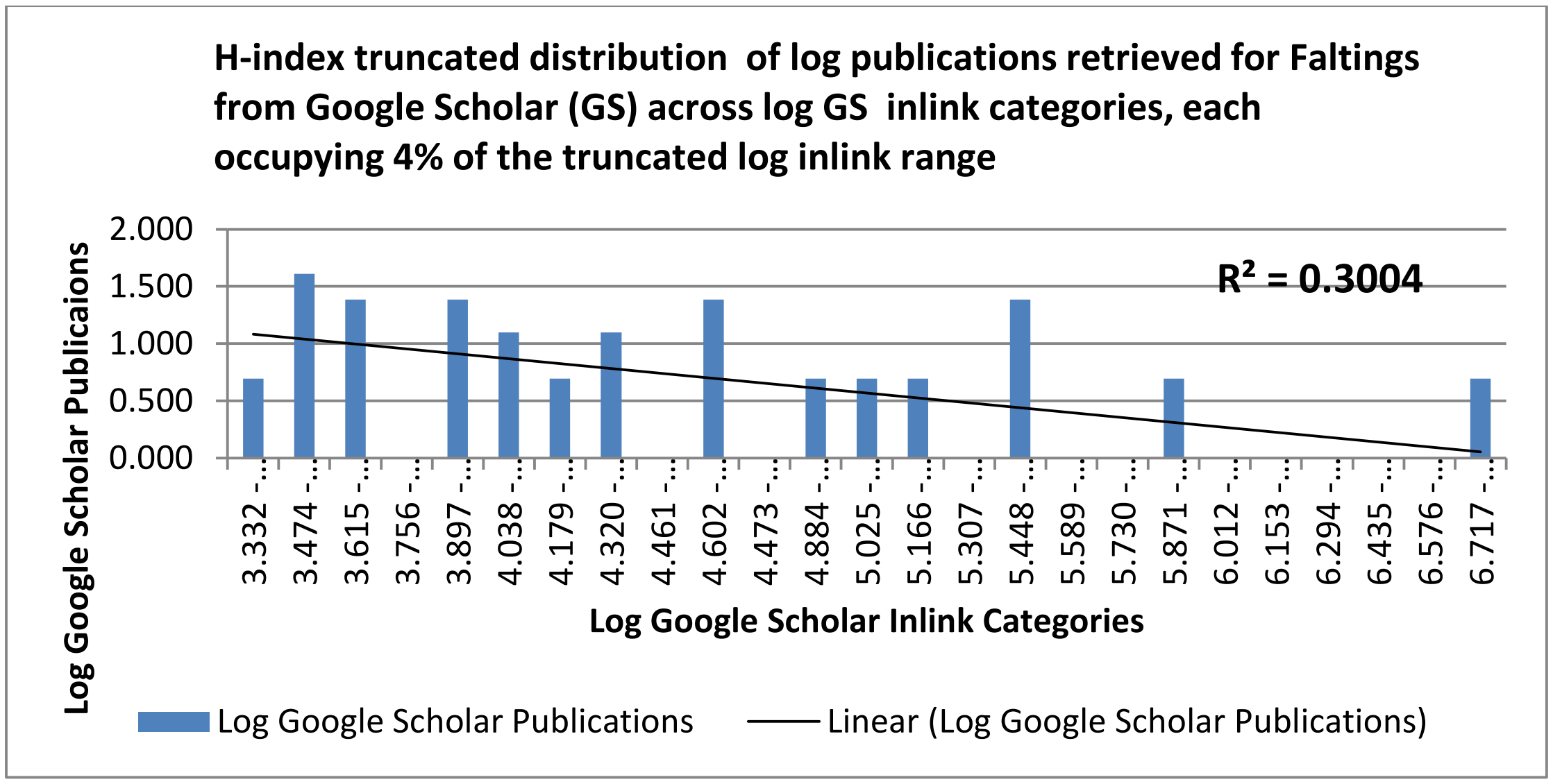

**F = 9.88. Fit is significant at the 1% level**

| Table 4. Linear fits (R^2) of log Google Scholar (GS) publications of Fields medalists to their log GS inlinks in the distribution truncated at the h-index. | | | |
|---|---|---|---|
| **Medalist** | **Year** | **R^2*** | **F**** |
| Simon K. Donaldson | 1986 | 0.1846 | 5.21 |
| Gerd Faltings | 1986 | 0.3004 | 9.88 |
| Maxim L. Kontsevich | 1998 | 0.3580 | 12.83 |
| Ngo Bao Chau | 2010 | 0.0072 | 0.17 |
| Elon Lindenstrauss | 2010 | 0.03797 | 0.91 |
| * The Excel LINEST function uses regression analysis to calculate the linear trendline, using the "least squares" method to find the straight line that best fits the data. R^2--or the coefficient of determination--runs from 0 or no fit of the line to the data to 1 or a perfect fit of the line to the data. | | | |
| ** For significance of the fit, LINEST not only produces the F statistic but also a degrees of freedom (df) statistic that enables calculating the v1 and v2 degrees of freedom necessary to enter published F-distribution tables at the proper place to determine the significance level. The formulas are v1 = n-df-1 and v2 = df, where n is the number of data points and df is the LINEST degrees of freedom. Our data was organized in such a way that n always equaled 25 and df always equaled 23. Therefore, v1 always equaled 1, and v2 always equaled 23. As a result, the same level of significance is applicable to all the tests above. To be significant, F has to 4.28 and above at the 5% level or 7.88 and above at the 1% level. | | | |

showing the distribution and the second showing the linear fit—indicates that the absence of linear fit was a result of the distribution not having formed. From this, it appears that there is no metric base for awarding the Fields medal. The R^2 of the earlier medalists do increase, reaching 0.358 with Kontsevich (1998) and 0.3004 with Faltings (1986). Both charts in Figure 7 for Faltings—one showing the distribution and the other the linear fit—manifest a resemblance to the equivalent ones for the most recent chemistry laureate, Negishi (2010), in Figure 3. However, the R^2 for the other of the earliest medalists, Donaldson (1986), is merely 0.1846 significant at the 0.05 level. These results sharply contrast those of the chemistry Nobelists, whose R^2 are all higher at much lower percentages of significance. Given the fragmented structure of mathematics, its slow citation rate, and the award of the Fields medal early in the careers of the researchers, these were the results that were expected. They validate not only the use of GS for assessing the publication impact of researchers (at least at the higher levels of the inlink distribution), but also confirm the employment of the Yule-Simon distribution as a mathematical construct able to measure how well research is being incorporated into the disciplinary knowledge corpus.

**Conclusion**

The application of the Yule-Simon model to GS data to measure the publication impact of winners of the Nobel prize in chemistry and the Fields medal in mathematics has delivered results that are consistent with the structural differences not only between the disciplines but also between the awards. This serves as validation for its use in assessing the researchers' publication impacts. Moreover, the findings of this chapter corroborate those of Egghe, Guns, and Rousseau (2011) in their study of the uncitedness of Nobelists and Fields medalists, which they stated thus:

> Contrary to what one might expect, Nobel laureates and Fields medalists have a rather large fraction (10% or more) of uncited publications. This is the case for (in total) 75 examined researchers from the fields of mathematics (Fields medalists), physics, chemistry, and physiology or medicine (Nobel laureates). (p. 1637)

The finding most "remarkable" to Egghe, Guns, and Rousseau was the positive correlation between the *h*-index and the number of uncited articles. They developed what they termed "a Lotkian model" (it can be considered "a Yule-Simon model") that partially explained the regularities found by them. As has been shown here, there is a tendency for the publications of such prize winners to concentrate in the lowest part of the inlink or citation range, and for mathematics, this is compounded by the fractured nature of the discipline. One of the advantages of the Yule-Simon model is that it is operative over time (i.e., given the proper conditions and enough time, the distribution will go to a limit, where the logging of both the independent and dependent variables results in a straight line with a negative slope). This aspect of the Yule-Simon distribution is not accidental, for Yule pioneered the time series analysis, or sets of ordered observations taken at different points in time (Hepple, 2005, pp. 990-991).

In this chapter, the Yule-Simon distribution was employed as a mathematical construct for testing (not describing) reality. Therefore, it was most informative when the observed distribution did not meet the criteria of the theoretical distribution, accordingly giving insights into what was actually taking place. This was certainly the case of the lack of any linear fit to the inlink distributions of the two 2010 Fields medalists—a result certainly in accordance with both the nature of mathematics as a discipline and the award itself. In this sense the Yule-Simon distribution was employed like the normal distribution, which is not descriptive of most reality but rather a mathematical construct to test for the amount of error in a data set. One advantage of the technique employed in this chapter is that it can be executed with Excel functions and is easily learned by statistical novices. The proof, however, which was presented in this chapter, cannot at all be considered definitive, because too many important facets of this problem could not be discussed. It is the intention of the authors to explore further this question together with others from the perspective of the theory of probability and citation indexing with special attention to how the Google Search engine winnows the wheat from the overwhelming chaff cluttering the web.  The latter problem was explored with the above sample of chemistry laureates by Bensman (2013), who proved that Google's PageRank algorithm is actually a further implementation of the theory of citation indexing of Eugene Garfield, creator of the *Science Citation Index*, at a higher technical level, yielding results fully in conformance Garfield's Law of Concentration, whereby the scientific journal system is dominated by a small multidisciplinary core of highly cited journals, and confirming his view of the importance of review articles.